\documentclass[a4paper,14pt]{extarticle}

\usepackage{graphicx}
\usepackage{amsmath,amsthm,amssymb}

\usepackage[bindingoffset=1cm,
            left=1.4cm,
            right=1.4cm,
            top=1.4cm,
            bottom=2cm,
            footskip=1cm]{geometry}
\usepackage{float}
\usepackage{enumitem}

\title{New ideas in nonperturbative QCD -- I\thanks{v2 -- 12/07/24}}
\author{ {M.S.~Lukashov}\thanks{m.s.lukashov@gmail.com},\quad {Yu.A.~Simonov}\thanks{simonov@itep.ru} \bigskip \\ NRC ``Kurchatov Institute'' \\ Moscow 123182, Russia}

\newcommand{\beq}{\begin{eqnarray}}
\newcommand{\eeq}{\end{eqnarray}}
\newcommand{\be}{\begin{equation}}
\newcommand{\ee}{\end{equation}}

\newcommand{\tr}{\operatorname{Tr}}

\def\fun#1#2{\lower3.6pt\vbox{\baselineskip0pt\lineskip.9pt
\ialign{$\mathsurround=0pt#1\hfil ##\hfil$\crcr#2\crcr\sim\crcr}}}

\newcommand{{\SD}}{\rm SD}

\newcommand{{\Mc}}{\mathcal{M}}

\newcommand{\vex}{\mbox{\boldmath${\rm x}$}}

\newcommand{\ver}{\mbox{\boldmath${\rm r}$}}

\newcommand{\veP}{\mbox{\boldmath${\rm P}$}}
\newcommand{\vep}{\mbox{\boldmath${\rm p}$}}
\newcommand{\veq}{\mbox{\boldmath${\rm q}$}}
\newcommand{\veQ}{\mbox{\boldmath${\rm Q}$}}

\newcommand{\vek}{\mbox{\boldmath${\rm k}$}}

\newcommand{\vev}{\mbox{\boldmath${\rm v}$}}

\newcommand{\lan}{\langle}
\newcommand{\ran}{\rangle}

\begin{document}
\maketitle
\begin{abstract}
The recent development of the Field Correlator Method (FCM) is discussed, with applications to the most interesting areas of QCD physics obtained in the lattice data and experiment. These areas include: \textit{a)} the connection of colorelectric confinement with the basic quark and gluon condensates; \textit{b)} the explicit form of the colorelectric deconfinement at a growing temperature $T$; \textit{c)} the theory of the colormagnetic confinement at all temperatures; \textit{d)} the theory of strong decays, the theory of pdf, and jets in the instantaneous formalism with confinement. We demonstrate that the FCM with instantaneous formalism and confinement (instead of the light cone formalism and pure perturbation theory) can provide the way to the theory of QCD, which helps to describe world data without phenomenological parameters.
\end{abstract}

\section{Introduction}

The phenomenon of confinement  creates more than 90 percent of the visible mass in the Universe and makes the world as we can see it. These years, QCD was feasting \cite{NRC,QCD50} the 70th anniversary of the first formulation of the simplest  QCD Lagrangian and 50 years of its final form \cite{1}, including the discovery of the asymptotic freedom phenomenon \cite{2}. At the present time, we have a well developed  theory of QCD  based on numerous experiments and lattice calculations, which allows to explain and predict phenomena in our world. Nevertheless, the existing theory is not yet complete, and in many areas, the standard theory has a descriptive character with fitting parameters (e.g., as in the standard theory of hadron spectra and thermodynamics). Another  difficulty with the standard approach to  QCD \cite{QCD50} is that it accepts a set of chosen models, such as QCD purely perturbation theory, chiral perturbation theory, string model, and others. However, all these models, when compared to experiment need phenomenological implications. The full QCD theory is impossible without a proper theory of confinement, which was started 35 years ago \cite{3} and is now internally consistent and able to explain not only the phenomenon of confinement  and all relevant data as will be discussed below, but also can serve as a building block for the whole QCD theory \cite{4}. The most important property of the confinement theory is that it also establishes the validity of other theories and phenomena in QCD, e.g., the perturbation theory with confinement acquires internal consistency both in the UV and IR regions \cite{5}, the string theory becomes an important section in the total comprehensive theory of hadrons in QCD \cite{6}, and the chiral theory with confinement becomes not a model but a full theory of the lowest mass mesons \cite{7}. Even the thermodynamics in QCD is not possible without confinement \cite{8} and deconfinement theory explaining the steep drop of string tension below $T_c$ \cite{9} and the colormagnetic confinement above $T_c$ \cite{10}.
The striking phenomena in the physics of hadrons and in QCD thermodynamics appear in the external magnetic field, which are fully explained by the existing confinement theory \cite{11}. This allows  the avoidance of the use of different qualitative models with extra parameters, which hardly makes a full, consistent theory.

One can find out that the theory of confinement is usually dropped in the standard theory textbooks where the property of confinement is treated  phenomenologically. At the same time the approach to the theory of confinement in the form of the Field Correlator Method (FCM) was suggested 35 years ago \cite{3} and was well confirmed by comparisons with recent lattice calculations \cite{4}. All the following years the development of FCM included the perturbation theory with confinement \cite{5}, exact formulation of the string theory, hadron spectra and decays \cite{6}, the chiral theory with confinement \cite{7}, theory of thermodynamics with confinement \cite{8}, and finally two important explanations of the temperature dependent confinement: 1) the colorelectric deconfinement -- the strong temperature behavior of the temporal string tension $\sigma_E(T)$ in the deconfining phase transition \cite{9}, and 2) the growth of the colormagnetic confinement with temperature \cite{10}. Especially clear confirmation of the FCM theory is the confinement in the magnetic field \cite{11} which is well supported by lattice data.

Recently a new important basic result has been obtained in \cite{12}: the fundamental scale of confinement was found  to be directly connected with the vacuum QCD energy -- the general  fundamental scale of QCD -- the gluon condensate $G_2$ \cite{13}. Indeed in \cite{12} the string tension $\sigma_E$ considered previously as an independent parameter was shown to be directly connected with the gluon condensate $G_2$. At the same time Standard QCD is faced with the challenge of explaining many new phenomena appearing in experiment and on the lattice, e.g. the high energy hadron-hadron cross sections \cite{14*,14}, the behavior of the hadron form factors \cite{15*,15}, quark condensate \cite{16*,16}, and etc. An important feature of the high energy processes can be connected to the instantaneous form of the strong dynamics \cite{17} suggesting the instantaneous dynamics with confinement to explain the experimental energy dependence of decays and scattering processes, which might also explain the phenomena of the jet quenching and ridge. 

As a result Standard QCD results based on historically introduced generally accepted and proved methods of pure perturbation theory may partly need some improvements. The following sections of the paper are:
\begin{description}
\item[1.] Introduction.
\item[2.] The basic nonperturbative scale of QCD, the sum rule of the hadron pressure and the gluon condensate.
\item[3.] The theory of confinement in QCD at $T=0$ -- the string tension from the gluon condensate.
\item[4.] The colormagnetic confinement at large $T$.
\item[5.] The QCD  dynamics in the instantaneous form and the resulting Green's functions and cross sections. Instantaneous strong decays and jets,Lorentz boost effects in high $Q$ processes and form factors.
\item[6.] Conclusions.
\end{description}
In the concluding section we discuss the possible lines of further development and adjustment of nonperturbative QCD.

\section{The basic scale of QCD}

The perturbative theory of QCD with the property of asymptotic freedom was historically  the basic part of QCD since the beginning and the perturbative basic length  $(\Lambda_{QCD})^{-1}$ remains the main characteristics of the QCD systems in Standard QCD discussions. It will be the main purpose for us below to consider the nonperturbative basic elements of QCD  defined by the phenomenon of confinement -- as it is in the external world around us where a more fundamental and nonperturbative  scale -- the vacuum condensate -- defines the basic dynamics \cite{12,13}.
 The most important step in the understanding of the nontrivial vacuum structure in QCD was done in
\cite{13} where the important role of the vacuum gluon condensate $G_2 = \alpha_s/\pi \lan G^a_{\mu\nu}G^a_{\mu\nu} \ran$  was discovered and its approximate magnitude $0.012 GeV^4$ was estimated from the charmonium sum rules. It is now understood that the vacuum gluon condensate is the basic property of QCD which stipulates the appearance of the main nonperturbative properties of QCD such as the confinement  and the chiral symmetry breaking \cite{7}.
At this point one should stress that this vacuum value of $G_2$ provides the resulting confinement string tension $\sigma_E$ \cite{12} and the quark condensate $\lan \bar{q}q \ran$  which were recently derived numerically  in a good agreement with experimental values \cite{17}.
This is in contrast with the standard  approach  to start with the perturbative regime in QCD and to use
accordingly the value of $\Lambda_{QCD}$ as the basic scale. As a result one can distinguish two
different ways: 1) perturbative scale  approach, 2) nonperturbative scale approach. In the first case
one can try to express all dynamic values at zero and nonzero temperature $T$ via $\Lambda$ and $T$ as e.g. in the derivation of the trace anomaly \cite{18} the grand potential $\Omega$ and the pressure $\Omega/V$ are derived as $P= \lambda^4 f(\lambda/T)$ and as a result one obtains the trace anomaly relation \cite{18}
\be
\frac{\beta(\alpha_s)}{4\alpha_s} \lan GG \ran= -\left(4-T
\frac{d}{dT}\right) P_h(T). \label{1} \ee
It is clear that the hadron pressure $P_h(T)$ is weakly connected with the perturbative interactions in QCD and therefore this widely demonstrated relation can be only
a small part of the much larger nonperturbative contribution which is not present in eq.~(\ref{1}). Practically the hadron pressure for light hadrons can be written as
$P_h(T)= T^4 f(M/T)$ where the hadron mass $M$ can be expressed via the gluon condensate $G_2$ and does not depend on  $\Lambda$ in the leading order. As a result the r.h.s. of the eq.~(\ref{1}) in the approach 2) turns to zero, which means that the dynamics based on the basic vacuum gluon condensate is not possible in the purely perturbative regime and the trace anomaly refers to the purely perturbative sector not including the most of the pressure. Indeed on the resulting pictures of the pressure one cannot see any connection of the resulting curves  to the deconfinement temperature as it is seen in the lattice data \cite{19}.
   Let us now turn to the general situation when the nonperturbative sector is important.In this case the pressure is mostly nonperturbative and should enter the basic relations on the same footing as the gluon condensate $G_2$. Indeed in \cite{9} another relation was suggested connecting gluon and quark condensates and hadron pressure $P_h(T)$ in a simple form
\be
|F_1(T)|=  |\epsilon_{vac}(T)| + P_h(T),\quad \epsilon_{vac}= 1/2 \epsilon_g + \epsilon_q,\quad \epsilon_q= \sum_q m_q \langle \bar{q} q \rangle,
\label{2} \ee
where the gluon condensate enters as follows
\be
\epsilon_g(T)= -\frac{b \alpha_s \langle G^2 \rangle}{32\pi},\quad \langle G^2 \rangle=\langle G^a_{\mu\nu}G^a_{\mu\nu} \rangle.  \label{3} \ee

We have taken into account in the eq.~(\ref{2}) that only the colorelectric gluon condensate $\langle G_E^2 \rangle= 1/2 \langle G^2 \rangle$ enters with the pressure which is accounted for by the coefficient $1/2$ before $\epsilon_g(T)$. We also note that the quark condensate
term $\epsilon_q$ as found in \cite{16} contributes around $(10-15)$ percent of the total vacuum energy and disappears approximately at the same temperature $T_c$ and therefore below we disregard it in the first approximation.
In eq.~(\ref{2}) $P_h(T)$ is the hadron interacting gas pressure growing with the temperature while the gluon condensate $\langle G^2(T) \rangle$ (its colorelectric part $G^2_{CE}$) decreases, vanishing at $T=T_c$. We suggested at this point in \cite{9} that the sum $F_1(T)$ in eq.~(\ref{2}) is constant in the temperature basis which defines behavior of the vacuum condensates depending on the hadron pressure. At the same time as found in our FCM method \cite{4}  the colormagnetic part of the gluonic condensate develops independently and it grows at large T  as it is exactly supported by lattice data \cite{19}. In this way the colorelectric and colormagnetic d.o.f. are found to be disconnected (in the first approximation).

At this point one must define the behavior of the CE vacuum energy $\epsilon_g(T)$ as a function of the temperature which will
explain the properties of the hadron gas and its deconfinement transition. In what follows we impose the following condition
on the confining free energy which will be called The Vacuum Dominance Mechanism (VDM) where the hadronic pressure is growing with
temperature $T$ with the simultaneous decrease of the vacuum energy (CE gluon condensate), so that their sum is kept constant.
\be
|F_1(T)|= 1/2 |\epsilon(T)| + P_h(T)= 1/2|\epsilon(T=0)|. \label{4} \ee

As shown in \cite{12} and will be discussed below the relations (\ref{2})-(\ref{4}) are in a good agreement with lattice  data  and will be considered below as
an essential part of our analysis. As it is the gluon condensate $G_2= \frac{\alpha_s}{\pi} \lan FF \ran$ defines the energy scale of the
QCD and it is important that it consists of colorelectric and colormagnetic parts $G_2= G_2^E + G_2^M$ which are equal at $T=0$
but develop in the different way at $T>0$ and especially at $T>T_c$ \cite{12}.

\section{The theory of confinement in QCD -- string tension from the gluonic condensate}

As shown in \cite{3} and developed in \cite{4} the phenomenon of confinement is created by  the bilocal
colorelectric field correlator $D^E(z)$ which is the vacuum average of two colorelectric fields $E_i(x)=F_{i4}(x)$ at the distance $z$ from each other
\begin{gather}
g^2 D^{(2)}_{i4k4} (x-y) \equiv \frac{g^2}{N_c}\,\lan \tr (F_{i4} (x) \Phi(x,y) F_{k4} (y) \ran = \notag \\
= (\delta_{ik}) D^E(x-y) + \frac12 \left(\frac{\partial}{\partial x_i} [h_k + {\rm perm.}]\right) D_1^E (x-y), \label{5} \\
h_\lambda = x_\lambda -y_\lambda, \,\, (x-y)^2 = \sum^4_{\lambda=1} (x_\lambda-y_\lambda)^2. \notag
\end{gather}
Indeed the appearance of nonzero correlator $D^E(z)$ leads to the area law of the Wilson loop $W(C)= \exp \left[ -\sigma S_{min} \right]$
with the nonzero colorelectric (CE) $\sigma_E$ or colormagnetic (spatial) $\sigma_s$ for the time-like or space-like surfaces
$S_{min}$, e.g. for the time-like surface one has
\be
\sigma_E = \frac12 \int d^2z D^E(z).\label{6}\ee
One can see that the bilocal FC $D^E$ yield the confinement for the time-like surfaces, and one can show that higher order
correlators can also ensure confinement.
As it is explained in \cite{4} the higher multilocal correlators are also present in the QCD vacuum but contribute less than
10 percent into the string tension. As a result one obtains the so-called Casimir scaling law for the confinement string tensions in different $q \bar q$ representations \cite{20} which is well supported by the lattice data \cite{21}.

For  the charge  representation $D$ the quadratic correlator $D^E$ \ref{5} defines the interaction between static charges in the representation $D=3$,$8$, $6$,...
\be
V_D (R) = C_D \int^R_0 (R-w_1) dw_1
\int^\infty_0 dw_4 D^E  \left(\sqrt{w^2_1+ w^2_4}\right), \label{7}\ee
where $C_D = 2\frac{C_2(D)}{C_2(f)}$, and $C_2 (D)$  is the quadratic Casimir coefficient for the representation $D$. E.g. the ratio of $D^E,\sigma_E$ for the adjoint charges (e.g. gluons) over fundamental charges (quarks and antiquarks) is $C_2(adj)/C_2(fund)= 9/4$. This basic law of the Casimir scaling allows one to check the validity of different
schemes of confinement suggested previously in the literature.
From now on the selfconsistent program of the confinement mechanism in QCD can be explained as follows.\bigskip

\noindent 1) The string tension $\sigma_E$ is defined by the lowest field correlator $D^E(z)$ \cite{4,12}
\be
\sigma_E= 1/2 \int d^2z D^E(z). \label{8} \ee
\bigskip

\noindent 2) The correlator $D^E(z)$ can be expressed as the Green's function of two gluons (two propagating gluon lines)accompanied by the adjoint fixed Wilson line $ \Phi(x,y)$. Due to confinement -- all three lines are connected by three pieces of the fundamental film. This construction is called -- the ``two-gluelump Green's function'' \cite{22}. At the same time the correlator $D_1^E(z)$ is the
``one-gluon gluelump Green's function'' consisting of one gluon propagating line and adjoint Wilson line connected by the adjoint film. Coming back to the $D^E(z)$ one can express it via the two-gluon gluelump Green's function
\be
D^E(z)= \frac{g^4 (N^2_c- 1)}{2} G_{2glp}(x,y) \label{9} \ee
where $G_{2glp}$ can be written as a path integral over two gluon plus a fixed Wilson line trajectories interacting within themselves as shown in the Appendix of the \cite{22}. Asymptotically it can be written as
\be
G^{as}_{2glp}(x,y)= \sum_n |\Psi^{2glp}_n(0)|^2 \exp \left[-M^{2glp}_n |x-y|\right] \label{10} \ee
This expression is valid for $|x-y| \gg \rho,\,\,\rho \equiv (M^{2glp})^{-1} \approx 0.08$ fm. Note, that both $M^{2glp}$, $\Psi^{2glp}$ are defined by the same $\sigma_E$ and keeping the lowest eigenvalues in eq.~(\ref{8}) one obtains in \cite{12}
\be
G^{(2glp)} (x)  \approx 0.108 ~\sigma_E^2 e^{-M_0^{(2g)}|x|} \label{11} \ee
At small distances $D^E(z \approx 0)$ is close to $G_2=\frac{\alpha_s}{\pi} \lan FF \ran$  and  one can express the integral over $d^2z$ at $z < \rho$ as \cite{12}
\be
D(z)= -4 N_c \alpha_V^2(z) G_2 + N_c^2 \frac{\alpha_V^2(z)}{2\pi^2} D(z_{\rm max}) \ln^2\left(\frac{z_{\rm max} \sqrt{e}}{z}\right), \label{12} \ee
As a result one obtains the selfcoupled system of eqs.~(\ref{6}) and (\ref{7}) at $|x|> \rho$ which does not contain any parameters besides $\sigma_E$ and $\sigma_E$ is expressed via itself since both $\Psi^{2glp}_n(0)$ and $M^{2glp}_n$ are expressed via $\sigma_E$.\bigskip

\noindent 4) At the same time in the low $|x|$ region $D^E(z)$ has the dependence on the external parameter- the gluon condensate $G^2$, which can be seen in its definition (\ref{4}). The explicit behavior of $D^E(x)$ for $|x|< \rho$ was obtained in \cite{12}
and is shown in Fig.~1 of \cite{12} where one can see a narrow peak at $|x|\approx \rho$, while at $x=0$ one obtains
\be
D^E(x=0)= \frac{\pi^2}{18} G_2,\,\, G_2= \frac{\alpha_s}{\pi} <\lan 0|F^a_{\mu\nu}F^a_{\mu\nu}|0 \ran \label{13} \ee
\be
D(0)= 0.15 D(\lambda_0)= 0.15 64 \pi^2 \alpha_V^2 \sigma^2 \exp \left[-M_0 \lambda_0 \right].
\label{14} \ee
Combining the behavior of $D(z)$ both below and above the point $\rho$ one can
 determine the  value of the gluonic condensate $G_2$, defined by the relation $D(0)= \frac{\pi^2}{18} G_2$ , namely,

\be
G_2= 1.69 \sigma^2 \alpha_V ^2= 0.054 \alpha_V^2 {\rm GeV}^4.
\label{15}
\ee
This relation allows to find $G_2$ as the basic scale, corresponding to  measured string tension $\sigma_E$.
The resulting connection  of $\sigma_E$ and $G_2$ was established in \cite{12} to be
\be
\sigma_E= \sqrt{\frac{G^E_2}{1.69 \alpha_s^2}} \label{16} \ee
where $G^E_2=1/2\,G_2= \frac{\alpha_s}{2\pi} (F^a_{\mu\nu})^2$ and one can keep the temperature dependence on both sides of eq.~(\ref{9}) as it was found in \cite{9}.

Summarizing one obtains the confinement phenomenon $\sigma_E(T)$ due to the vacuum field correlator averages can be expressed
via the gluon condensate $G^E_2(T)$ defining the basic scale of the confining system.
In this way we have defined the connection of the confinement with the fundamental structure of the QCD vacuum at zero temperature where $G^E_2(T=0)=G^M_2(T=0)$ and respectively $\sigma_E(T=0)= \sigma_s(M)(T=0)$ as it follows from the general
definitions \cite{4} and is supported by the lattice data.
It is clear that at zero temperature all contours $C$ and the minimal surfaces inside  them in eq.~(\ref{5})are equivalent and therefore both temporal $\sigma_E(T=0)$ and spatial string tension $\sigma_s(T=0)$ coincide. However for $T>0$ they may differ
and as will be seen later above the deconfinement temperature $T_c$ the first vanishes while the second grows.
As was discussed above the behavior of the temporal string junction is dictated by the Vacuum Dominance Relation (VDR) (see eq.~(\ref{4})) which is supported by the lattice data  \cite{19}. Using eqs.~(\ref{4}) and (\ref{16}) one obtains the simple relation
\be
\sigma_E(T)= \sqrt{1 - \frac{P_h(T)}{P_h(T_c)}} \label{17} \ee
The comparison of the eq.~(\ref{17}) with the lattice data is given in Fig.~1 and one can see a reasonable agreement.

\begin{figure}[!htb]
\begin{center}
\includegraphics[width=0.8\linewidth]{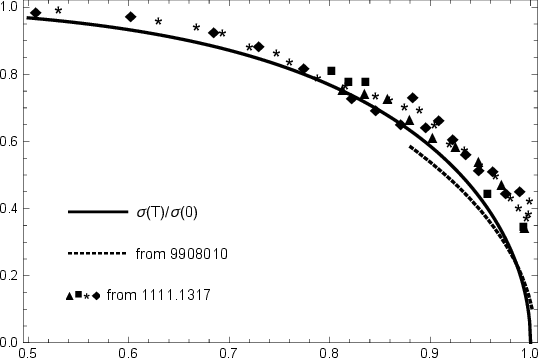}
\caption{Comparison of the lattice data for the ratio $\sigma(T)/\sigma(0)$ from \cite{19} -- dotted line and dots, with our result from eq.(\ref{15}) -- solid line.}
\end{center}
\label{Fig1}
\end{figure}

In this way one can see a strong drop of the temporal string tension at the deconfinement temporature
which is accompanied by the chiral symmetry restoration \cite{7}.

\section{The colormagnetic confinement at growing temperature T}

The behaviour of the spatial string tension $\sigma_s(T)$ as a function of the temperature $T$ was found in the framework of the Field Correlator Method (FCM) in \cite{23,24,25}. Here the string tension is calculated using the gluelump Green's function \cite{22,23}, where gluons in the gluelump are interacting via the same spatial string tension interaction. The resulting selfconsistent T-dependence was obtained in \cite{10} without extra parameters in the region$T_c < T < 5 T_c$ using the formalism of elliptic  functions $\theta_3$.
The string tensions $\sigma_E,\sigma_s$ are defined
 via the vacuum field correlators  of the
colorelectric (CE) and the colormagnetic (CM) fields $E_i^a,H_i^a$, which differ in the light-like areas $\sigma_E$ and space-like areas
$\sigma_H$  but coincides at the zero temperature $T$,
$\sigma_E(T=0)=\sigma_H(T=0)= \sigma$.
The CE and CM field correlators are defined as bilocal vacuum averages of the CE and CM fields
\be
D^{(E,H)}(x-y)= \frac{1}{N_c} \left\langle \tr \left\{(E_i,H_i)(x)\Phi(x,y)(E_i,H_i)(y) \right\} \right\rangle.  \label{18} \ee
The field correlators $D^E(x)$, $D^H(x)$ define all confining QCD dynamics and in particular the string tensions and $\Phi(x,y)$ is the Wilson line that  connects points $x$ and $y$.
\be
\sigma_E= \frac12 \int (d^2z)_{i4} D^{E}(z),\,\,\sigma_H= \frac12 \int (d^2z)_{ik} D^{H}(z). \label{19} \ee

It was found in \cite{26,27} that the dominant part of the spatial string tension $\sigma_s(T)$ grows quadratically at large $T$
\be
\sigma_s(T)= (c_{\sigma})^2 g^4(T) T^2,  \label{20} \ee

where $c_{\sigma}$ was defined numerically in the lattice calculations \cite{26,27} in the case  $N_c=3$, $N_f=0$ as
\be
c_{\sigma}= 0.566\pm 0.013.  \label{21} \ee

 On the theoretical side the quadratic growth of the $\sigma_s(T)$ was derived in the framework of FCM
 \cite{24,25} and the  value of $c_{\sigma}$ was found in \cite{24}
 in a good agreement with the lattice data of \cite{26,27}.

However in the full FCM expression for the spatial string tension
the term in eq.~(\ref{18}) is only a fast growing part  of the whole expression which was hitherto not known.
As one can see in eqs.~(\ref{16}) and (\ref{17}) the field correlator $D^E(x,y)$, $D^H(x,y)$ in the nonabelian case due to the
relation e.g $F_{\mu\nu}= \partial_{\mu}A_{\nu}- \partial_{\nu}A_{\mu} - ig [A_{\mu}A_{\nu}]$ can be represented as a sum of terms where  one or two gluons propagate along the Wilson line $\Phi(x,y)$ interacting between themselves and with the Wilson line via confinement (with $\sigma_{E,s}$) interaction -- this construction is called below ``one- or two- gluon gluelump'' following \cite{22,23} which should reproduce the string tension ($\sigma_E$ or $\sigma_s$) in a selfconsistent way as was explained in \cite{10} yielding the internal structure of the nonperturbative  QCD vacuum. As was found in \cite{4,24,25} the string tension
is defined by the two-gluon gluelump propagator with two gluons and the Wilson line all connected by adjoint strings with the string tension $\sigma_a= \frac94 \sigma_f$. In this way the calculation of the string tension is a selfconsistent process which we describe below for the spatial string tension.

The spatial string tension is proportional to the integral of this two-gluon  gluelump Green's function in the $3d$ space, where one of three space coordinates can be taken as an evolution parameter (the Euclidean ``time"). Using the technic, developed in \cite{22,23} for
$D^E(z)$, $D^H(z)$, which allows to express it via the two-gluon  Green's function: $G^{(2g)}_{4d} (z) =  G_{4d}^{(g)} \otimes G_{4d}^{(g)}$, where two gluons and the Wilson line (as it is called ``the parallel transporter'')
interact nonperturbatively, and  we  neglect  the spin interactions in the first approximation.
At this point we omit the detailed derivation of the string tension given in \cite{10,24,25} and  we turn to the general form of the field correlator $D^H(z)$ with the aim to express the string tension via the calculable factors $f(x)$. One has
\be
D^H(z) = \frac{g^{4}(T)(N^2_c-1) }{2} \lan G^{(2g)}(z,T) \ran,
\label{22}
\ee
where $G^{(2g)}(z,T)$ is the gluelump Green's function
\be
G^{(2g)}(z,T)= \frac{z}{8\pi} \int \frac{d\omega_1}{\omega_1^{3/2}} \frac{d\omega_2}{\omega_2^{3/2}} D^{3}r_1 D^{3}r_2
\exp \left[ (-K_1-K_2-V(\ver_1,\ver_2)z) \right].
\label{23}
\ee
As a result one obtains $\sigma_s(T)$ in the following form
\begin{gather}
\sigma_s(T)= \frac{g^{4}(T)(N_c^2-1)}{4} \int d^2z z/(8\pi) \int d\omega_1 d\omega_2 (\omega_1\omega_2)^{-3/2} \times \notag \\
\times \sum_{n=0,1,} |\psi_n(0,0)|^2 \exp \left[ -M_n(\omega_1,\omega_2)z \right] f(\sqrt{z/2\omega_1}T)f(\sqrt{z/2\omega_2}T).
\label{24}
\end{gather}
Here the function $f(x)$ is defined as follows
\be
f(x)= \sum\limits^{+\infty}_{n=-\infty}e^{-n^2/(4x^2)}
\label{25} \ee

\noindent The integrals in eq.~(\ref{22}) without factors $f(cT)$ do not contain the temperature dependent factors, and
one can see in eq.~(\ref{24}) the only T-dependent factors $g^4(T)$ and $f(\sqrt{z/2\omega_1}T)$ which define the dependence of $\sigma_s(T)$. Therefore one can write $\sigma_s(T)$ (denoting the $z$- and $\omega$- integration in eq.~(\ref{24}) with the average sign
$\lan \ldots \ran$) in the following form
\be
\sigma_s(T)={\rm const}\,g^4(T) \left\langle f^2(\sqrt{z/(2\omega)}T) \right\rangle.
\label{26} \ee

We can consider the average value of $\sqrt{z/2\omega}$ (obtained as a result of integration over the $T$-independent region of parameters with the $T$-independent kernel) as  a fixed ($T$-independent) parameter to be
  checked by the comparison with lattice data .
  The appearance of $g^4(T)$ which is decreasing with $T$ as $(\ln T)^{-2}$ defines the $T$
  dependence of $\sigma_s(T)$ to be lower than $T^2$, thus confirming the behaviour of $\sigma_s(T)$
  in the lattice data of \cite{26,27}, where the data were fitted
   as $\sigma_s(T)= {\rm const} g^4(T) T^2$ . However this fit fails for $T<2T_c$
   claiming the necessity of another factor in eq.~(\ref{24}).
   Correspondingly we are writing the resulting equation for the $\sigma_s(T)$ denoting the average value of
   $\sqrt{z/(2\omega)}T$ as a constant (T-independent) parameter  which we denote as $\rho T/T_c$, where $\rho$ is a number. As a result one obtains the equation for the string tension
\be
\sigma_s(T)={\rm const}\,{g^4(T)} f^2(\rho T/T_c) .
\label{27}
\ee

Using eqs.~(\ref{25})-(\ref{27}) one can write
$f(w)= \sum \limits^{+\infty}_{n=-\infty}\exp\left[-\frac{n^2}{w^2}\right]\equiv\vartheta_3(q)$, $q=\exp\left[-\frac{1}{w^2}\right]$
with $w^2=\frac{\rho^2 T^2}{T_c^2}$. Correspondingly the $f(<\lan w \ran)$ acquires the form
\be
f(<\lan w \ran) = F(T/T_c)= \vartheta_3 \left( \exp\left[{-\frac{T_c^2}{(\rho T)^2}}\right] \right)
\label{28} \ee
 The numerical analysis of the data \cite{19} allows to
reproduce well the data with the equation of the form \be
\sigma_s(T)=\sigma_s(T_c) \frac{g^4(T) F^2(T/T_c)}{g^4(T_c)F^2(1)}
\label{29} \ee

The analysis of the lattice data in comparison with eq.~(\ref{27}) is shown in Fig.~2, where for $g^4(T)$
the explicit value of the $L_\sigma= 0.104 $ as in Ref. \cite{19} was used while in $f(<\lan w \ran)$ in eq.~(\ref{28})
the value $\rho=3$. Fig.~2 demonstrates  good agreement between the lattice data and  eq.~(\ref{27}),
including the region $ T< 2.5 T_c$ where the lattice fit $T^2 g^4(T)$ starts to disagree with numerical data.

\begin{figure}
\center{\includegraphics[width=0.7\linewidth]{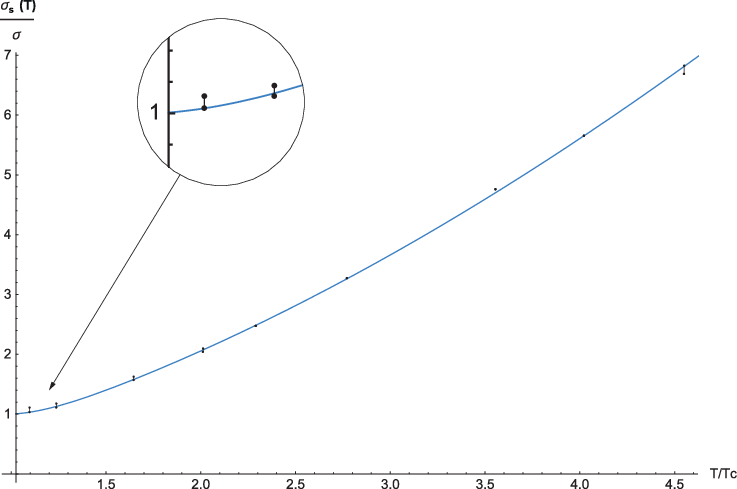}}
    \caption{Spatial string tension $\sigma_s (T)/\sigma$ for SU(3)
gauge theory as function of $T/T_c$.  The lattice data with errors are from Ref. \cite{26}, $T_{c}$=270 MeV.}
    \label{Fig2}
\end{figure}
One can see a remarkable agreement between the theoretical and experimental data.
 
\section{Light cone or instantaneous interaction in QCD}

 As it is known, \cite{17} in the  relativistic  field theory the general formalism can be constructed in three different ways:
\begin{description}
\item[1)] the instant form,
\item[2)] the point form, and
\item[3)] the light front form.
\end{description}
For the discussion in the literature see e.g. \cite{28}.
The perturbative QCD was the first item in the development of QCD but its role in the dynamics of the world is subsidiary and can be reduced to the perturbative corrections to the basic QCD processes at all temperatures. Therefore in what follows we concentrate mainly on the nonperturbative phenomena of QCD such as confinement and chiral symmetry breaking. Another important issue of the general QCD dynamics: is it of the light-cone type,as it is exploited in numerous papers now, or it is of the instantaneous type \cite1{17} as it is natural for the confinement interaction. As it was suggested before \cite{28} and  appears now ,supported by recent studies \cite{29,30,31} the observable strong interactions proceed via the instantaneous type of interactions which  may naturally explain the decrease of the decay rate \cite{29} and  the high $Q$ effects \cite{30,31}
This important topic is connected with the processes at high energy and the high momentum transfer -- as it is discussed in \cite{32}.
At present it is popular to establish the light-cone (LC) dynamics which is convenient for the  formulation of the perturbative effects.However the hadron spectra in the LC formalism do not reproduce well the experimental spectra \cite{33,34,35} and therefore the LC formalism should be abandoned in the exact nonperturbative calculations and will not be discussed below. The main topics of this section are connected with the instantanwous form (IF) of the strong  dynamics in QCD and we shall argue that the main features
of the observed dynamical effects in QCD can be explained in the framework of the IF dynamics.
The main points of this discussion below can be listed as follows:
\begin{description}
\item[A)] the hadronic decay amplitudes in the IF dynamics vs experiment,
\item[B)] the IF effects in the hadron form factors,
\item[C)] the new form of the high energy parton distributions  via Lorentz boosted hadron wave functions and finally,
\item[D)] the multihybrid  halo of high energy collisions as a source of jet quenching and ridge phenomena.
\end{description}

\subsection{(A) The hadron decay width in the instantaneous decay dynamics}
  In the instant form the wave function of any nonlocal object consisting of several elements can be defined at one moment of time and the frame (boost) dependence is dynamically generated in connection with Hamiltonian.  The instant form is especially convenient to introduce the confinement in the system of several quarks and antiquarks
  when the confinement strings couple together all constituents. Namely this formalism was exploited in the applications  of the FCM \cite{3,4,5,6} to the calculations of the hadron spectra, correlators and decays  in  good agreement with lattice and experiment.The same is true for the decay matrix elements where both initial and final
hadron wave functions are taken at the same moment of time. However in this approach one should take into account that when  the decaying hadron and decay products are moving with high  velocities and hence in the rest system of the decaying hadron  the wave functions of the decay products are modified by their velocities and in this case one should take into account the Lorentz transformation of the the wave functions of the moving hadrons \cite{29} .The same is true
  for any matrix elements containing  wave functions of moving hadrons, e.g. for the hadron form factors where hadron is moving with different velocities before and after collision with the photon \cite{30,31}.
   As it is, the theory of the frame dependence of the Green's functions of any nonlocal objects  is closely related to the properties of the interaction terms in the Lagrangian, and one must envisage the instantaneous interaction for the first formalism, in particular confinement for the strong interaction and the Coulomb force in QED. The dynamical studies in this direction have been done
 recently, in Refs.~\cite{29,30,31} in several examples of systems. Later on the properties of the spectrum and the wave functions in the moving system were studied in the framework of  the relativistic path integral formalism \cite{32}.  This method essentially exploits the universality and the Lorentz invariance of the Wilson-loop form of interaction, which produces both confinement and the gluon-exchange interaction in QCD. Moreover, in this formalism the Hamiltonian $H$ with the instantaneous interaction between quarks in QCD (called the relativistic string Hamiltonian (RSH)) and charged particles in QED was derived  and therefore the known defects of the Bethe-Salpeter approach are missing there. In the instant form (IF) approach it was shown in \cite{29,30,31,32} that the eigenvalues and the wave functions, defined by the RSH, transform in the moving system in accordance with the Lorentz rules. Indeed, using the invariance law under the Lorenz transformations one has
\be
\rho(\vex,t)dV = {\rm invariant},\label{30}
\ee
where $\rho(\vex,t)$ is the density for the wave function $\psi_n(\vex,t)$,
\be
\rho_n(\vex,t) = \frac{1}{2i} \left(\psi_n \frac{\partial \psi_n^+}{\partial t}  - \psi_n^+ \frac{\partial \psi_n}{\partial t}\right)
= E_n |\psi_n(\vex,t)|^2,\label{31}
\ee
and $dV=d\vex_{\bot} dx_{\|}$. Using the standard Lorentz transformations one has
\be
L_{\rm P}dx_{\|} \rightarrow dx_{\|} \sqrt{1 - \vev^2}, ~~ L_{\rm P} E_n \rightarrow \frac{E_n}{\sqrt{1-\vev^2}},
\label{32}
\ee
to insure the invariance of eq.~(\ref{28}). As a result the wave function $\psi(\vex,t)=\exp\left[-i E_n t \right]\varphi_n(\vex)$ the function $\varphi_n(\vex)$ is deformed in the moving system,
\be
L_{\rm P}\varphi_n(\vex_\bot, x_{\|}) = \varphi_n\left(\vex_\bot, \frac{x_{\|}}{\sqrt{1-\vev^2}}\right),
\label{33}
\ee
and can be normalized as
\be
\int E_n |\varphi_n^{(v)}(\vex)|^2 dV_v = 1 = \int M_0^{(0)} |\varphi_n^{(0)}(\vex)|^2 dV_0,
\label{34}
\ee
where the subscripts $(v)$ and $(0)$ refer to the moving and the rest frames. One of the immediate consequences from the
eqs.~(\ref{32}) and (\ref{33}) is the property of the boosted Fourier component of the wave function:
\be
\varphi_n^{(v)}(\veq) = \int \varphi_n^{(v)}(\ver) \exp(i\veq\ver) d\ver = C_0 \varphi_n^{(0)}(\veq_\bot, q_{\|}\sqrt{1-v^2}),
\label{35}
\ee
where $C_0 = \sqrt{1-v^2} = \frac{M_0}{\sqrt{M_0^2 + \veP^2}}$.

The equations~(\ref{30})-(\ref{31}), and in particular eq.~(\ref{31}), are  the basic elements of the analysis of the meson form factors in \cite{30}, where it was shown that the Lorentz contraction
of the hadron wave functions creates a basically different behavior of form factors as functions of $Q^2$, such that arguments of wave functions are never in the asymptotically large region of momenta. In the concrete examples of the
 pion and kaon form factors the agreement with data was obtained with simple Gaussian wave functions in the whole region of $Q^2$. A similar situation holds for the proton and neutron form factors .

We start with the simplest form of the $^3P_0$ model with the interaction Hamiltonian
 \be H_I = g \int(d^3x \bar\psi \psi), \label{36} \ee
 where $g = 2 m_q \gamma$, and $\gamma$ is a phenomenological parameter. The relativistic form obtained in \cite{29}
 can be written as
 \be S_{eff} = \int{ d^4 x \bar\psi(x) M(x) \psi(x)}, \label{37} \ee
 where $M(x) = \sigma (|\vex -\vex_Q| + |\vex - \vex_{\bar Q}|)$. Here $\vex$ is string breaking point between the quarks $Q$ and $\bar Q$.
In the momentum space one obtains as in \cite{17,18} for the decay of the hadron $1$ into hadrons $2,3$
\be J_{123}(\vep) = y_{123} \int{\frac{d^3q}{2\pi^3} \Psi_1(\vep , \veq) M(q) \psi_2(\veq) \psi_3(\veq)}.
\label{38} \ee

Here $\vep$, $\vep$ are the momenta of decay products and $\veq$ are the internal momenta inside decay products, which we assumed to be identical for simplicity.

Moreover $y_{123}$ is the trace of normalized spin-tensors corresponding to spin-angular parts of meson states and $M(q)$ for the $S$-wave decay is a constant, proportional to the string tension, $M(q)= \mathcal{O}$($1$ GeV) and for the $L$-wave resonance it is proportional to the $p^L$. Finally for the width one can write

\be \Gamma(E)= {\rm const} \sim p(E)^{2L + 1} |J(p(E))|^2. \label{39} \ee
Here $L$ is the angular momentum of the decay products.
So far we are in the realm of the standard hadron decay formalism. We now take into account that the decay product wave
functions are moving with the velocity $\sqrt {\frac{s - (m_1 + m_2)^2}{s}}$, and hence their wave functions in momentum space are  Lorentz contracted as shown in eq.~(\ref{35}).
To this end we must write $J(p(E))$ in terms of the contracted wave functions, namely as in eq.~(\ref{6}), the wave function moving with the velocity $v$ can be written as $\psi_n^{(v)}(\veq) = C_0 \psi_n(\veq_\bot, q_{\|}\sqrt{1-v^2})$. Denoting the total energy $E$ which coincides with the resonance mass at the
resonance center,as $s=E^2$, one can write $C_0 = \sqrt{1-v^2}= \frac{m_2 + m_3}{\sqrt{s}}$.
Therefore the integral in eq.~(\ref{39}) can be rewritten as
\begin{gather}
J(p) = {\rm const} \int(d^3\veq\Psi_1^{0}(\veq , \vep)\psi_2^{v}(\veq)\psi_3^{v}(\veq)) = \notag \\
= {\rm const} \sim C_0^2 \int(d^2\veq_\bot dq_{\|}\Psi_1^{0}\psi_2(\veq_\bot,q_{\|}\sqrt{1-v^2})\psi_3(\veq_\bot,q_{\|}\sqrt{1-v^2}))= \notag \\
= {\rm const} \sim C_0 \int(d^2\veq_\bot d\kappa \Psi_1^{0} \psi_2(\veq_\bot,\kappa)\psi_3(\veq_\bot,\kappa)). \label{40} \end{gather}

Here $\kappa = q_{\|}\sqrt{1-v^2}$. Therefore the decay matrix element is multiplied by $C_0$ and the decay width is multiplied by $C_0^2$. Summarizing one can write for the two-body decay width of a resonance with account of Lorentz contraction , which we write first in the case of equal masses $m_2 = m_3 = m$

\be \Gamma({\rm LorC}) = C_0^2 \Gamma(0) = \frac{4 m^2}{s} \Gamma(0). \label{41} \ee

Here $\Gamma(0)$ denotes the decay  width without LorC dynamics.
Now one can consider the $\rho(770)$ meson as an example of a P-wave meson decaying into two pions with a large energy release and therefore subject to strong LorC corrections.

In PDG  it is written: `` ... the determination of the parameters of the $\rho(770)$ is beset with many difficulties because of its large width. In physical region fits, the line shape does not correspond to a relativistic Breit--Wigner function with a P-wave width, but requires some additional shape parameter.''
Indeed in the standard theory with the Lagrangian $L_{eff}= g_{\rho\pi\pi} e_{ijk}\rho^{i\mu}\pi_j \partial_\mu \pi_k$ one obtains the width
   \be \Gamma_\rho = \frac{g_{\rho\pi\pi}^2 p^3}{48\pi m_\rho^2}, p= \sqrt{s-4 m_\pi^2}, \label{42} \ee
the result which contradicts experimental data.
The numerous accurate experimental data exploit the corrected equation for the
$\Gamma_\rho (s)$, namely

\be \Gamma_V (s) = \frac{m_V^2}{s} \frac{p(s)^3}{m_V^3} \Gamma. \label{43} \ee

This should be compared with our result in eq.~(\ref{36}) for the decay of the $\rho \rightarrow \pi\pi$, where $\Gamma(0)$
refers to the width without LC, which is proportional to $p(s)^3$ and hence two equations coincide up to the replacement
of $4 m^2$ by $m_V^2$, which is unimportant since $\Gamma_V$ is a variable numeric parameter. In this way we have shown the importance of the LorC correction effects in
the energetic hadron decays.

\subsection{(B) Hadron form factors in the IF dynamics}

The equations~(\ref{28})-(\ref{33}), and in particular eq.~(\ref{33}), are  the basic elements of the analysis of the meson form factors in \cite{30}, where it was shown that the Lorentz contraction
of the hadron wave functions creates a basically different behavior of form factors as functions of $Q^2$, such that arguments of wave functions are never in the asymptotically large region of momenta. In the concrete examples of the
 pion and kaon form factors the agreement with data was obtained with simple Gaussian wave functions in the whole region of $Q^2$. A similar situation holds for the proton and neutron form factors \cite{33}.

As a result the hadron form factor in the Breit frame acquires the form \cite{30}
\begin{gather}
F(Q^2) = \int \varphi_{-\frac{Q}{2}}(\vek)\varphi_{Q/2}\left(\vek+\veQ\frac{\omega_2}{\omega_1 + \omega_2}\right) d^3k = \notag \\
= C_0^2(Q) \int \varphi_0(\vek_{\bot}, k_{\|}\sqrt{1-v^2}) \times \notag \\
\times\varphi_0\left(\vek_{\bot},\left(k_{\|} + Q \frac{\omega_2}{\omega_1+\omega_2}\right)
\sqrt{1-v^2}\right) \frac{d^3k}{(2\pi)^3}.
\label{44}
\end{gather}
Here $C_0(Q)=\frac{M_0}{\sqrt{M_0^2 + Q^2/4}}$. The comparison in \cite{30} of the $\pi$ and $K$ experimental formfactors with the theory  shows a good agreement.
The further analysis of the asymptotics of the hadron form factors in \cite{15} allows to obtain the well known relations -- the ``quark counting  rules''
\be
F_h(Q_0)=\left(\frac{4 M_h^2}{Q_0^2}\right)^n_h,\,\,Q_0^2= 4 (M_h^2 + vep^2), \label{45} \ee
Here $h$ refers to mesons with $n_M=1$ and baryons with $n_B=2$.
We can compare the experimental data for the pion form factor with our pedictions from eq.~(\ref{44}) as it is done below

\begin{table}[!htb]
\caption{Comparison of the experimental pion form factor \cite{37} with theoretical predictions from eq.~(\ref{4})}
\begin{center}
\label{tab01}
\begin{tabular}{|l|c|c|c|c|c|}
\hline &&&&&\\
$Q^2$ (in GeV$^2$)            & 0.6         &    0.75           &  1.0          & 1.6         &  2.45\\\hline
&&&&&\\

$m_\pi$ (in GeV)                      &  0.20         &   0.191             &   0.181     &    0.168     &   0.166  \\\hline
&&&&&\\
$F_\pi\,(\exp.)$ \cite{37,39}   & $0.433$ & 0.341 &0.312   &$ 0.233$ & $0.167$ \\
\hline
&&&&&\\
$F_\pi\,({\rm mod.,\,th.})  $        &    0.43     &     0.375          &   0.316         & 0.238         &    0.188 \\

\hline
\end{tabular}
\end{center}
\end{table}
In Tab.~\ref{tab01} one can see a reasonable agreement between theoretical and experimental values within $O(25\%)$ accuracy and with the accuracy better 10\% for $Q^2\geq 1.6 \sim $ GeV$^2$. At the same time the asymptotic behavior, at  $Q^2\geq 1.6 \sim $ GeV$^2$, is given with a good accuracy and does not imply standard perturbative behavior \cite{38}.

We now turn to the $K^+$ meson form factor treating in the same way as above for the pion case (see \cite{32}).     
\begin{table}[!htb]
\caption{Comparison of the calculated function $Q^2F_K(Q^2)$ for the $K^+$ meson form factor ($k_K=0.23$~GeV) with experimental data \cite{39} and the lattice data \cite{40} \cite{41}}
\begin{center}
\label{tab02}
\begin{tabular}{|l|c|c|c|c|c|}
\hline
&&&&& \\
$Q^2$                           & 0.10    &  0.5   &   1.0           & 1.5     &  2.5    \\ \hline
&&&&& \\
$Q^2F_K(Q^2)$ (th.)             & 0.0874  &  0.28  &  0.38           &  0.44   &  0.48  \\ \hline
&&&&& \\
$Q^2F_K(Q^2)$ (exp.)     &        &        & $0.37\pm 0.12$  &         & $0.45\pm 0.04$ \\ \hline
&&&&& \\
$Q^2F_K(Q^2)$ (lat.)    & 0.08     &  0.28  &  0.38          &          & 0.48  \\
\hline
\end{tabular}
\end{center}
\end{table}
One can see in Tabs.~\ref{tab01} and \ref{tab02} a reasonable agreement of the Lorentz contracted forms with data which supports the use of the instantaneous formalism (IF) in the hadron processes.

\subsection{(C) Parton distributions via Lorentz contracted hadron wave functions}

At this point we are entering a dangerous region of a basic disagreement between the purely perturbative and the nonperurbative mechanisms of the wave functions in  QCD. In the first case the free relativistic quarks and gluons are assumed to interact perturbatively with the evolution mechanism given by the DGLAP equations \cite{38}. In this case one meets with the IR and collinear singularities which are treated with the corresponding cutoffs. Recently a new nonperturbative approach based on the instantaneous Lorentz contracted forms was developed
for the quark-gluon evolution at high energy and momentum in \cite{31,32,42} which allows to study the whole Fock component structure of the stable systems and the equivalent quark-gluon systems in the parton model (``parton distributions from Fock components'').
The total wave function of quarks and gluons $\Psi_N$ can be decomposed into a sum of Fock components as follows \cite{42}
\be
\Psi_N= \sum\limits_{m(k)} c^N_{m(k)} \psi_{m(k)};\,\,\psi_{n(k)} = \psi_{n(P,\xi,k)}\label{46}\ee
Here $n(k)$ refers to number and types of constituents etc.
 The quark and gluon momenta should be decomposed into longitudinal and transverse parts
 \be
 x_i= \frac{p_{||i}}{P},\,\,P\sqrt{1-v^2}= M_0   \label{47}  \ee

  As a result the parton distribution function (pdf) in the hadron  can be written as follows \cite{42}

\be
D^q_h(x,k_{\perp})= \frac{M_0^2}{(2\pi)^3} \left| \phi(k_{\perp},M_0(x-1/2)) \right|^2, \label{48}  \ee
Here $\phi(\vek)$ is the meson wave function normalized as $\frac{M_0}{(2\pi)^3} \int {d^3 k}|\phi(\vek)|^2 = 1$
In this way the lowest pdf is the strongly accelerated ground state w.f. It has a maximum at $x=1/2$ as in the standard
pdf but no singularity at $x=0$. The latter and the full contents of the pdf appears when one takes into account the
not only the basic wave function as in eq.~(\ref{8}) but also higher Fock components  $\phi_N$ \cite{31,42,44}.
In a similar way one obtains the pdf of the nucleon starting from the nonperturbative wave function of the proton. E.g. for the $u$-quark projection of the wave function one has (see eq.~(10) from the second Ref. of \cite{32})
\begin{gather}
u\left(x, k_{\perp}\right)=\int \delta^{(2)}\left(\sum_{i=1}^3 k_{\perp i}\right) \prod_{i=1}^3 d^2 k_{\perp i} d x_1 d x_2 d x_3 \delta\left(1-\sum x_i\right) \times \notag \\
\times \frac{M_0^3}{(2 \pi)^3}\left|\tilde{\varphi}_0^{(3)}\right|^2\left[\left(\delta^{(2)}\left(k_{\perp}-k_{\perp 1}\right) \delta\left(x-x_1\right)+(1 \leftrightarrow 2)\right)\right] \label{49}
\end{gather}
where $\tilde{\varphi}_0^{(3)}$ is $\tilde{\varphi}_0^{(3)}\left(k_{\perp 1}, \ldots k_{\| 1}^{(0)}, \ldots\right)$.
 
Now using the hyperspherical formalism \cite{45} for the calculation of the nonperturbative QCD proton wave function obtained with account of the confinement between 3 quarks brings about the u quark pdf in Fig.~3  which can be compared to
the model and experimental data. Indeed this is shown in Fig.~3 taken from \cite{32},where the result from eq.~(\ref{45})  is compared with the standard model results from \cite{46,47}. One can see a maximum of  $u(x)$ around $x=1/3$ which is reproduced both by the DGLAP-type functions as well as the pure 3q hyperspherical wave function.
In this way we demonstrate that the nonperturbative nucleon wave function is able to account for the main features of the parton distributions opening in this way the possibility of to formulate the full nonperturbative parton theory. It is instructive that the hadron wave functions obtained with the help of confinement and other nonperturbative effects produces the same type of the pdf which is produced by the purely perturbative methods and considered as a series support of the standard (no-confinement) theory.

\subsection{(D) Multihybrid states at hogh energy and momentum-jet quenching and ridge  effect}

Interesting phenomena are observed in the high energy hadron-hadron and nucleus-nucleus collisions where the emitted hadrons can form some collective sequences with the specific features. These can be classified as follows:
\begin{enumerate}
\item suppression of the emitted high momentum hadrons,
\item formation of the hadron-emission sequences called jets \cite{41}, and
\item the formation of the correlated back-to-back hadron assemblies -- ridge phenomenon \cite{48}.
\end{enumerate}
As one can understand the numerous attempts to understand these phenomena in the framework of the Standard QCD based on  the dynamics of the perturbative QCD does not imply the creation of long lines of moderately energetic decays. Below we shall consider the nonperturbative and basic dynamical effects which are capable of creating the long almost straight lines of subsequent strong decays. We shall exploit two basic mechanisms:
\begin{enumerate}
\item the instant form (IF) dynamics discussed in this section above,
\item the possible formation of the multihybrid states \cite{42} in the h-h or A-A collisions.
\end{enumerate}
We start with the mechanism A) and write the decay distribution in the IF from eq.~(\ref{29}), where it was found that the decay width in the IF is proportional to the factor
$ (C_0)^2$ where $C_0= \frac{2m}{E}$ and as a result $\Gamma(E)= \frac{4m^2 \Gamma(2m)}{E^2}$
In a similar way the meson form factor is modified by the Lorentz contraction in the IF as follows
(assuming that the hadron wave function is gaussian and the hadron mass is $m_h$)
$$ F(Q^2)= \frac{m_h}{\sqrt{m_h^2 + Q^2/4}} \exp\left[ -\frac{Q^2 m_h^2}{16 k_0^2(Q^2/4 +m_h^2)} \right].$$
One can see a strong reduction of the asymptotic decrease of the form factor at the large $Q$ which is
 a consequence of the wave function flattening -- the basic feature of the Lorentz contracted hadron wave function. Finally it is interesting what happens with the multiple subsequent decays of the products of high energy collisions. In the IF strong dynamics with confinement one observes the creation of the multiple products consisting of multihybrids with many gluons which subsequently decay into ordinary
 hadrons \cite{42}. This creates a good condition for the formation of the jet quenching phenomena \cite{44} since the Lorentz contraction of the wave functions in the decay matrix elements stipulates
 progressive decays with small energy (forming the jets) and for decays with large energy products must be emitted at large angle-this imitates jets sequences interrupted by high $Q$ deflections observed in experiment \cite{44}.
 In this way we see a completely different picture of the high energy collisions, decays and form factors based on the constantly present confinement, Lorentz contraction and high excited multihybrid
 states in the high energy collisions.
In this way we can identify two basic effects of the nonperturbative interactions in the high energy $pp$, $AA$ collisions: domination of the small energy release in the high excited hadron decays and creation of the excited objects-multihybrids-which decay consequtively with relatively small energy release $\mathcal{O}(1\text{\ GeV})$ at each step. These nonperturbative effects combine with basic perturbative jet mechanisms stressing the main features of the jet quenching phenomenon -- long lines of decays with relatiely
low energy release and appearance of the back-to-back jet trajectories -the ridge effect \cite{48}.
We start with the phenomenon of the successive decays of an excited hadronic object with the restricted energy release. As was discussed above there are at least two
mechanisms which can strongly limit the energy release: a) the Lorentz contraction of the fast decay products \cite{29} and b) possible formation of the multihybrids \cite{42}.
In the first mechanism discussed in the previous subsection, eqs.~(\ref{37}) and (\ref{38}), the decay width into several hadrons is proportional to the product $(1-v_1^2)(1-v_2^2)\ldots$ where $v_i$ is a velocity of the $i$-th decay product. This effect as was discussed above strongly reduces the width of the $rho$ meson and its excitations.
The effect b) can be of the crucial importance since the formation of the multihybrids in the high energy $h$-$h$ or $A$-$A$ collisions can be a direct consequence of the pomeron
and odderon exchanges between colliding hadrons \cite{49} -- since the pomeron is a bound state of two gluons interacting via confinement and gluon exchanges and the $s$-channel
discontinuity of these exchange diagrams naturally dissects the multihybrid construction. Therefore it is interesting to get more info about the enrgy distribution in the multihybrids and to compare it with the energetic scales in jets. In multihybrids with $n$ gluons the total mass according to \cite{42} is
$M_n \approx 1.24$ GeV n .Transforming the multihybrid results into the PDF form one obtains in \cite{42} for the gluon distribution $xg(x)$ at $x= 10^{-2}$, $10^{-1}$ the
values $3.87$;$1.38$ which is close to the PDG values at $Q^2= 10\text{ GeV}^2$ namely  $5$; $1.5$. (for more similarities between the standard QCD and the multihybrid physics see eq.~(\ref{42})). Summarizing this subsection one can see in the phenomenon of the jet quenching additional features of the Lorentz contraction in IF dynamics and the multihybrid formation in high energy collisions,which call for more intensive analysis of these phenomena outside of the standard perturbation theory.

\section{Conclusions}

We have described above a small piece of a new possible picture of the strong interaction theory in QCD, based on the ever present confinement in the QCD dynamics at zero temperature and in high energy and momentum processes. Our outlook is a bit unconventional and different from the common perturbative picture basically without confinement (when the latter is introduced as a secondary factor or a small correction), which is present in the Standard QCD approach.

This new picture is only starting to develop and needs additional checks and applications in all areas
of QCD, including qgp plasma physics, theory of nuclear matter and nuclei, theory of neutron stars and etc., where the role of the confinement and  the correct dynamical formalism necessary for the confinement is not yet taken into account. Actually the instantaneous dynamics (in the c.o.m. static
system) was used successfully in the derivation of the Green's functions and relativistic Hamiltonian in the FCM approach in \cite{4,5,6,7,8} and exploited for the treatment of the spectra and processes. Correspondingly all relativistic calculations of the spectra and pomeron and Regge trajecories within the FCM were done using the instantaneous Hamiltonian in the static system \cite{36}. However the dynamics of the decay transitions and high-energy scattering and the hadron emission requires an explicit account of the Lorentz contraction for the hadron wave functions, which was proposed and studied in \cite{29,31}.

Summarizing we have illustrated the proposed complete dynamical scheme for the calculation of different processes in QCD which fully takes into account confinement both in the hadron structure and in the hadron transitions and decays. At the same time we propose the fundamental relations of the confinement dynamics with the basic vacuum structure-the quark-gluon condensate. This allows to define the explicit form of the colorelectric deconfinement and the colormagnetic growth with the increasing temperature.

\end{document}